\documentclass{article}

\usepackage{neurips_2024}
\usepackage{tcolorbox}
\usepackage{enumitem} 

\usepackage{graphicx}
\usepackage[utf8]{inputenc} 
\usepackage[T1]{fontenc}    
\usepackage{hyperref}       
\usepackage{url}            
\usepackage{booktabs}       
\usepackage{amsfonts}       
\usepackage{nicefrac}       
\usepackage{microtype}      
\usepackage{xcolor}         

\usepackage{listings}

\title{Multi-Agent System for Cosmological Parameter Analysis}

\begin{document}

\maketitle

\begin{abstract}
 Multi-agent systems (MAS) utilizing multiple Large Language Model (LLM) agents with Retrieval Augmented Generation and that can execute code locally may become beneficial in cosmological data analysis. Here, we illustrate a first small step towards  AI-assisted analyses and a glimpse of the potential of MAS to automate and optimize scientific workflows in Cosmology. The system architecture of our example package, that builds upon the \texttt{autogen/ag2}\footnote{\href{https://github.com/ag2ai/ag2}{https://github.com/ag2ai/ag2}} framework, can be applied to MAS in any area of quantitative scientific research. The particular task we apply our methods to is the cosmological parameter analysis of the Atacama Cosmology Telescope lensing power spectrum likelihood using Monte Carlo Markov Chains. Our work-in-progress code is open source and available at \href{https://github.com/CMBAgents/cmbagent}{https://github.com/CMBAgents/cmbagent}.
\end{abstract}

\section{Introduction}

In many disciplines, scientific discoveries are now driven by the sheer volume of datasets and our ability to interpret them. Familiarizing ourselves with these datasets, data formats, and analysis pipelines can be time-consuming and painstaking. With recent progress in natural language processing and information retrieval, LLMs offer an unprecedented opportunity to streamline and optimize these processes. But what is the right way to integrate LLMs into scientific workflows?

\begin{figure}
        \centering        \includegraphics[width=0.8\linewidth]{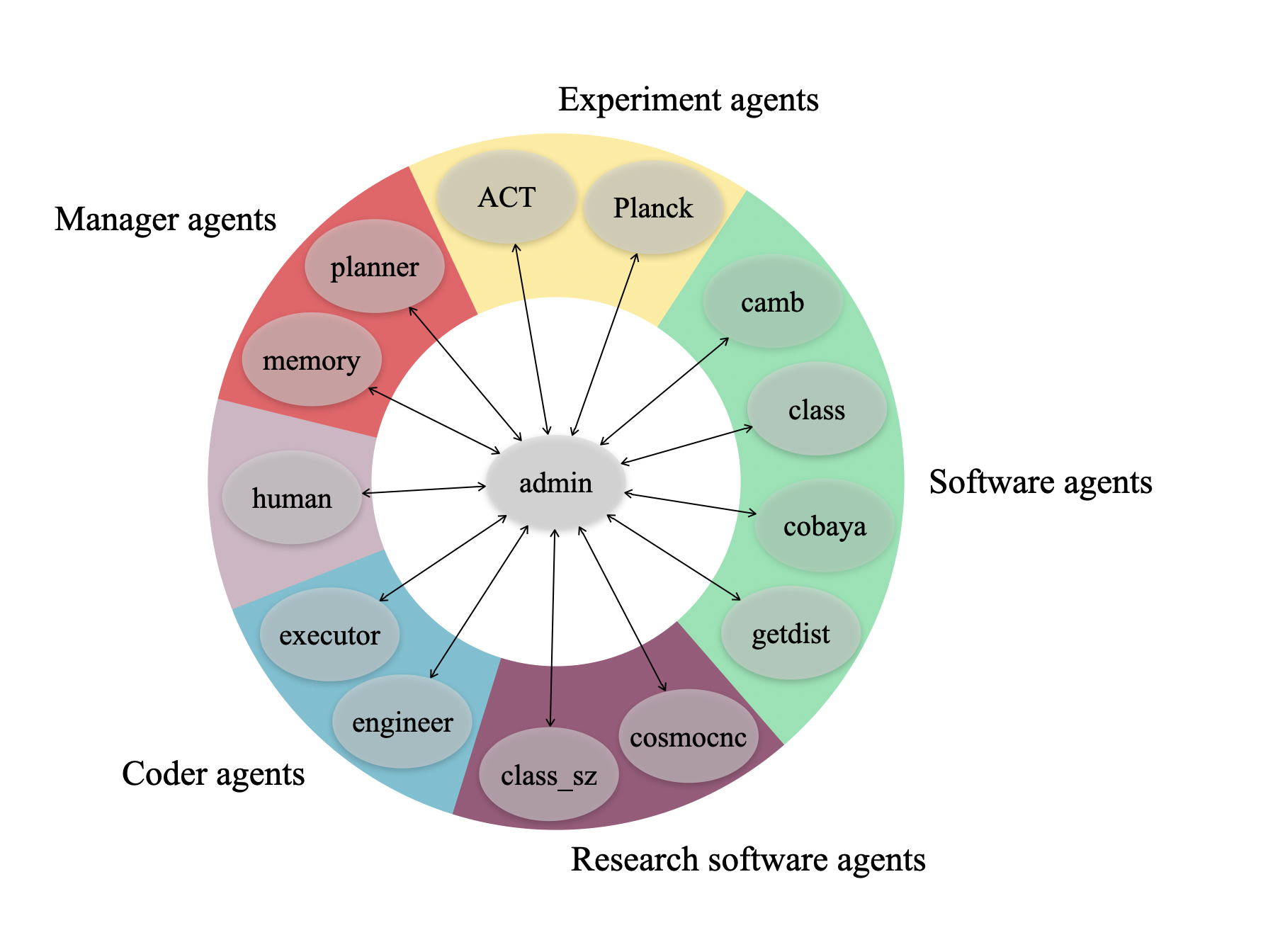}
        \caption{Overview of system hierarchy and transitions. Manager agents are shown in salmon, coder agents are shown in blue, experiment RAG agents in yellow, community software RAG agents in light green, and research software RAG agents in magenta. Arrows indicate current allowed transitions in \href{https://github.com/CMBAgents/cmbagent}{\texttt{cmbagent}}.}.
        \label{fig:AutogenDiag}
    \end{figure}

Here, we explore the agentic avenue.
By employing multiple specialized LLM agents, we can break down complex scientific tasks into manageable sub-tasks, each handled by an expert agent. These agents collaborate to retrieve information, write and execute code, and manage transitions between sub-tasks until they reach their designated common objective. As an implementation of this approach, we focus on state-of-the-art cosmological data analysis tasks, typically carried out within large scientific collaborations. Nevertheless, we emphasize to our strategy can be applied to any area of research in the physical sciences involving quantitative data analysis.

We implement our multi-agent system (MAS) with the open source \href{https://github.com/ag2ai/ag2}{\texttt{autogen}} programming framework \citep{wu2023autogen}.\footnote{Originally \texttt{autogen} was created within Microsoft and we used the first implementation for the examples presented here, i.e., \href{https://github.com/microsoft/autogen}{\texttt{microsoft/autogen}} The code now lives in a separate organization called \href{https://github.com/ag2ai/ag2}{\texttt{ag2ai/ag2}}, which is going to be used for the next iteration of \texttt{cmbagent}.} Although \texttt{autogen} allows for programming LLM agents with different language models, such as Llama \citep{2023arXiv230213971T}, OpenAI Generative Pre-trained Transformers (GPT) \citep{2020arXiv200514165B, 2023arXiv230308774O},\footnote{See, e.g., \cite{2023arXiv231012069T} for an introduction on the transformer architecture aimed at astrophysicists.} and Anthropic's Claude, in the cases presented here we focused solely on GPT-4 \citep{2023arXiv230308774O}. More specifically, we use  gpt-4o-2024-05-13 queried via the OpenAI Application Programming Interface (API).

Past studies have examined the use of MAS in different scientific applications, for example, \texttt{sciagents} for materials discovery \citep{sciagents,buehler2024graphreasoning}. In the realm of astronomy, at around the same time as \texttt{cmbagent} was developed, \cite{mephisto} developed a framework known as \texttt{mephisto} that used MAS to interpret multi-band galaxy observations, with similar aims as ours.\footnote{We note that our work-in-progress was presented in July at \href{https://indico.ict.inaf.it/event/2690/timetable/?print=1&view=standard_numbered}{ML4ASTRO2}, and our code is publicly available on GitHub since September 7, 2024.} A key aspect of our MAS is that it uses etrieval-augmented generation (RAG) (see Sec.~\ref{sec.method}).

Beyond the initial investigations presented in this manuscript, our broader aim is to address the following two questions: 
\begin{enumerate}[label=(\roman*)]
    \item Can we automate state-of-the-art cosmological data analysis pipelines in a generic way and that can later be extended to other scientific domains?
    \item Can a MAS find more optimal pipeline solutions than those made by humans?
\end{enumerate}

\section{Method}
\label{sec.method}

Here, we focus on (i), the first of the questions above. To build our agentic system example, called \texttt{cmbagent}, we first pick a data analysis task that is complex, uses current software, involves cutting edge data, and to which we already know a solution. Based on this task, we construct the MAS so that it can be generalized to other tasks not considered during its construction. That means that none of the codes and instructions we implement should refer to specific details of the main task chosen. We then assess the usefulness of the MAS when we apply it to new tasks.

\subsection{Main task}

The main task we chose to focus on while designing our MAS is the following: 

\begin{tcolorbox}[colback=gray!10!white, colframe=gray!70!black, title=Main Task]
Derive cosmological parameter constraints from ACT DR6 CMB lensing data
\end{tcolorbox}

This task involves Atacama Cosmology Telescope (ACT) Data Release 6 (DR6) Cosmic Microwave Background (CMB) lensing data, collected and maintained by the ACT Collaboration. 

The relevant ACT data was released and described in two articles \citep{ACT:2023kun, ACT:2023dou, maccrann2023atacamacosmologytelescopemitigating} along with a public likelihood code available online. Based on this, we can implement the likelihood and perform parameter inference as done by the ACT Collaboration. 

Cosmological parameter constraints obtained by the ACT Collaboration were based on Bayesian inference, using Markov Chain Monte Carlo (MCMC) sampling. To predict the signal, which is summarized into the lensing convergence power spectrum, the ACT Collaboration used \texttt{camb}. 

The \texttt{camb} code is a Boltzmann solver written in Fortran that computes cosmological perturbations across cosmic time and can predict summary statistics of the CMB \citep{Lewis:1999bs}. Along with its C language counter-part, \texttt{class} \citep{class}, it is one of the most widely used codes in cosmology. 

For the MCMC sampling the ACT Collaboration used \texttt{cobaya} \citep{Torrado:2020dgo}, and for the kernel density estimation (going from samples to posterior probability distribution), it used \texttt{GetDist} \citep{Lewis:2019xzd}. 

Over the past decade, these software packages (i.e., \texttt{camb}, \texttt{class}, \texttt{cobaya}, \texttt{getdist}) have been used in thousands of cosmology articles confronting theory and observations. They are part of the core curriculum that most cosmology Ph.D. students have to master and can be qualified as \textit{community software}. These software packages are all handled by our example MAS.  

As researchers, we develop our own software packages dedicated to our active area of research. Such software is often not well documented and used by a small number of expert users. Such software can qualified as \textit{research software}. 

A feature of the MAS that we are building which we want to explore is whether it can be capable of running calculations and write pipelines involving such research software. Here we consider two such pieces of software: \texttt{classy\_sz} \citep{Bolliet:2023eob} and \texttt{cosmocnc} \citep{Zubeldia:2024zow}.

\texttt{classy\_sz} is a machine-learning accelerated CMB and Large Scale Structure code written in Python and C, that builds on top of the \texttt{class} infrastructure. It is highly parallelized and uses deep neural network emulators for the matter power spectrum. Together, this makes the model evaluation optimally fast. The emulators are made with TensorFlow \citep{2016arXiv160508695A} and \texttt{cosmopower} \citep{Spurio_Mancini_2022}  \citep[see][for details]{Bolliet:2023sst}. 

\texttt{cosmocnc} is a Python package for computing the number-count likelihood of galaxy cluster catalogs in a fast, flexible and accurate way, enabling cosmological inference with state-of-the-art and upcoming galaxy cluster data. It is based on the use of Fast Fourier Transform convolutions in order to efficiently evaluate some of the integrals in the likelihood, and its core theoretical input, the halo mass function, is computed in a fast way with the \texttt{cosmopower} neural networks.

In cosmology, as in any other scientific discipline, the core curriculum of a researcher also includes knowledge of key results and literature associated with recent data and  experiments. In CMB physics, this would include literature and results from the ACT and \textit{Planck} Collaborations. The two primary methods used to specialize LLMs to a specific field or context are fine-tuning and Retrieval Augmented Generation (RAG) \citep[see, e.g.,][]{RAG}. Here, we focus on the latter. This motivates the introduction of agents of three different types that we describe hereafter.

\subsection{Agent types}

\textbf{(i) RAG agents:} RAG agents are specialized in information retrieval on experiments, community software, and research software. They come in two primary categories, \textit{experiment} and \textit{software} agents. They are OpenAI assistants with the \texttt{file\_search} functionality enabled. These agents have access to databases containing the relevant information which has been vectorized using OpenAI \texttt{text-embedding-3-large} model. Experiment agents' contextual data mainly consists of papers describing the respective experiments, such as \cite{ACT:2023dou} and \cite{ACT:2023kun} for ACT, or \cite{Planck:2018vyg,Carron:2022eyg} and \cite{Tristram:2023haj} for \textit{Planck}. Meanwhile, software agents' contextual data mainly consist of tutorial notebooks that explain how the software can be used. (Contextual data for all RAG agents is stored locally, within our GitHub organization,\footnote{See under \href{https://github.com/CMBAgents/cmbagent_data}{\texttt{cmbagent\_data}} online.} and is vectorized and pushed online upon installation or upon request.) RAG agents are shown in green (community software), magenta (research software), and yellow (experiments) on Figure \ref{fig:AutogenDiag}. 

In addition to the experiment and software RAG agents, there is a \textit{memory} RAG agent. The goal of the memory agent is to learn from mistakes and their solutions in past tasks, and apply them to future similar tasks. The memory agent also allows speed-ups by drawing on past material, and saves in cost by reducing the number of calls to other RAG agents. At the end of every session, the user has the option to upload a chat summary of their task to their memory agent's database, adding to the pool of information to which the memory agent has access in the future.

\textbf{(ii) Coder agents:} Coder agents are responsible for coding tasks. Following \texttt{autogen}, we use an \textit{engineer} agent to develop the code and an \textit{executor} agent to execute the code. We program them to be solely specialized in Python. The engineer agent is tasked with implementing all suggestions and information retrieved by the RAG agents. Its instructions are constraining it to edit one Python code block. It also knows about recurrent conflicts between settings or parameters that appear frequently, such as setting two cosmological parameter values that should not appear together because of redundancy (for example, the cosmological parameters $A_s$ and $\sigma_8$). In most situations, there is a back and forth between engineer and admin/human, until the code block written by the engineer is validated. Once the code is validated, admin/human sends instruction for the executor to execute the code. This is the only task of the executor agent. Its main input is the working directory (which by default is the output folder within \texttt{cmbagent}) and the code block edited by the engineer agent. If the outcome of the execution is satisfying, admin/human can call planner to summarize the session and save the summary with the help of the engineer and executor agents. If the outcome is not satisfying, a back and forth between admin/human and engineer, and potentially the RAG agents may resume to edit new corrected code from agents' suggestions. Coder agents are shown in blue on Figure \ref{fig:AutogenDiag}.

\textbf{(iii) Manager agents:} Manager agents are designed to distribute the work and organize feedback. We have a \textit{chat manager} agent responsible for  selecting the next relevant ``speaker" in the workflow, i.e., which agent should be called next, given a task and a set of allowed transitions. (In fact, the \textit{chat manager} agent implicitly uses a two-agent nested chat to decide the next speaker.) To pass on human feedback to the MAS, we use an \textit{admin} agent which collects human input in an interactive text box. To split the main task into a number of sub-tasks we build a \textit{planner} agent. Our system also includes a memory agent that performs RAG on summaries of past sessions, generated by the planner agent. Manager agents are shown in salmon on Figure \ref{fig:AutogenDiag}.

\subsection{System and workflow}

Because of the nature of the scientific applications for which we want to use \texttt{cmbagent}, we demand that our system is as deterministic and controllable as possible. Current large language models are generally stochastic. To minimize stochasticity we set both the \texttt{temperature} and \texttt{TopP} parameters of GPT-4o to small values (we chose $10^{-6}$ and 0.1, respectively). We implement the following solutions for improved controllability.

\textbf{Allowed transitions:} As an input to each agent, we specify the agents to which it can speak. This is a rich subject, especially when investigating MAS without human input. Aiming for maximum control,  we require human feedback at all stages. Allowed transitions are therefore trivial, consisting of admin  $\leftrightarrow$ $\{$Planner, RAG agents, Engineer, Executor$\}$, i.e., ``admin speaks to all, all only speak to admin''. 

\textbf{Planning:} Having a planner agent in the workflow is a key aspect of controllability. The planner agent speaks first. It designs a plan by splitting the main task into a number of steps consisting of one sub-task per step. In each step, the planner decides and specifies which agent should be selected. Importantly, only one agent can be in charge of a given sub-task. The admin may request modifications to the suggested plan until the plan is deemed ready. Hereafter is an example of a plan obtained when solving our main task.

\begin{tcolorbox}[colback=gray!10!white, colframe=gray!70!black, title=\textbf{Plan example}]
\textbf{Main task:} Derive cosmological parameter constraints from ACT DR6 CMB lensing data.
\begin{itemize}
    \item \textbf{Step 1:}
    \begin{itemize}
        \item \textbf{sub-task:} Retrieve information on how to set up the ACT DR6 CMB lensing likelihood within cobaya.
        \item \textbf{agent:} \texttt{act\_agent}
    \end{itemize}
    \item \textbf{Step 2:}
    \begin{itemize}
        \item \textbf{sub-task:} Retrieve information on how to use \texttt{classy\_sz} as the theory code within \texttt{cobaya}.
        \item \textbf{agent:} \texttt{classy\_sz\_agent}
    \end{itemize}
    \item \textbf{Step 3:}
    \begin{itemize}
        \item \textbf{sub-task:}  Integrate the ACT DR6 CMB lensing likelihood and \texttt{classy\_sz} theory code within the \texttt{cobaya} framework.
        \item \textbf{agent:} \texttt{cobaya\_agent}
    \end{itemize}
    \item \textbf{Step 4:}
    \begin{itemize}
        \item \textbf{sub-task:} Verify the entire setup, including the ACT DR6 CMB lensing likelihood and \texttt{classy\_sz} theory code within \texttt{cobaya}.
        \item \textbf{agent:} \texttt{engineer}
    \end{itemize}
    \item \textbf{Step 5:}
    \begin{itemize}
        \item \textbf{sub-task:} Execute the entire setup to ensure it runs correctly and derive cosmological parameter constraints.
        \item \textbf{agent:} \texttt{executor}
    \end{itemize}
\end{itemize}
\end{tcolorbox}

The full session that led to this plan is the documentation online (see \href{https://cmbagent.readthedocs.io/en/latest/notebooks/cmbagent.html}{this URL}). Designing a plan involves a back and forth between admin and planner agent. The instructions given to the planner agent are also available in \href{https://github.com/CMBAgents/cmbagent/blob/main/cmbagent/planner/planner.yaml}{\texttt{cmbagent/planner/planner.yaml}}. As for the instructions to the other agents (also available online on our GitHub in the yaml files for each agent), finding instructions that yielded a good behavior for our tasks required a significant amount of prompt engineering.

\textbf{Workflow:} When the plan is approved, the chat manager identifies which agent speaks next to work on the sub-task corresponding to the step in the plan. The RAG software agents suggest codes, RAG experiment agents collect information, engineer agent writes code, and executor executes code. The admin is asked for feedback after each suggestion or output from agent. Typically, the admin (i.e., human user) would say ``proceed'' when the response is acceptable, or ask for explicit modifications if not (see the examples \href{https://cmbagent.readthedocs.io/en/latest/notebooks/cmbagent.html#Cosmology-session}{here}). A back and forth can occur between the admin and the agent assigned to the task, until the sub-task is considered completed which triggers a move to the next step in the plan. The workflow generally ends with a call to the engineer agent followed by the executor agent. The engineer agent considers the entire chat history (consisting of the suggestions from all agents previously called and solutions to all sub-tasks). With this information, the engineer agent edits the python script that solves the main task entirely. The admin agent checks the scripts and eventually asks executor to run the code. 

\textbf{Learning from past tasks:} At the end of the process, the planner can be asked to provide a summary of what has been done and issues encountered. The user has the option to save the summary using formatted output as a \texttt{json} file. This file is saved into the database of the memory agent. In subsequent runs, the memory agent is queried to detect if previous similar tasks have already been solved and if so, which errors can be avoided.

\begin{figure}
        \centering
        \includegraphics[width=0.8\linewidth]{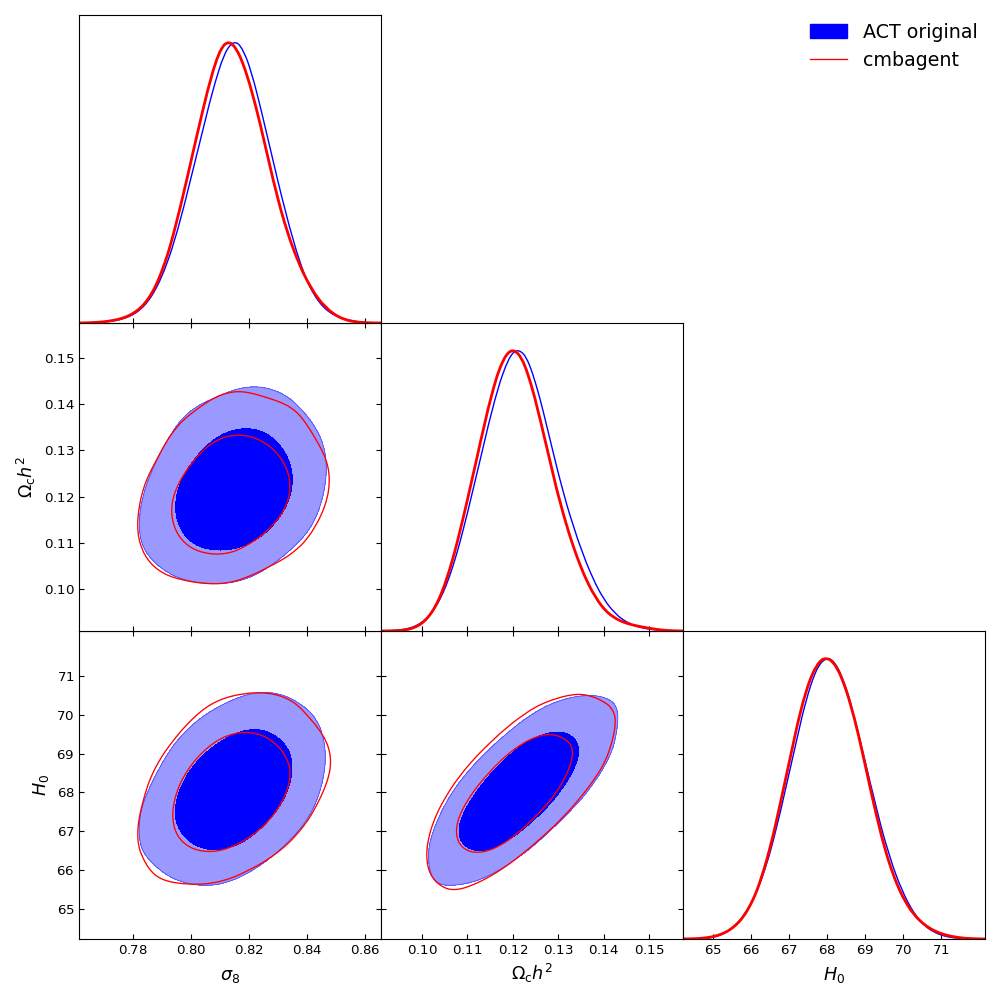}
        \caption{Reproducing of pipelines for the ACT DR6 CMB lensing cosmological parameter constraints. Contours obtained by our system, \texttt{cmbagent} are in red and the original chains downloaded from ACT repository are in blue. These results are obtained in the examples presented in the online documentation.}
        \label{fig:contours}
\end{figure}

\section{Results}\label{sec:limitations}

\subsection{Performance on Main Task} 

To develop \texttt{cmbagent}, we assigned ourselves the task of reproducing the ACT DR6 lensing cosmological parameter constraints. In Fig. \ref{fig:contours} we show our resulting constraints, obtained within a single session of our MAS. These results can be reproduced from within the notebook provided as part of our online documentation (see \href{https://cmbagent.readthedocs.io/en/latest/notebooks/cmbagent.html#Cosmology-session}{this URL}).

 As well as writing the code, our MAS also ran the full MCMC analysis by executing the code. The speedy evaluation of the theory model is enabled by neural network emulators built into \texttt{classy\_sz}. The full agentic analysis was carried out on a Macbook Pro laptop. The MCMC used 4 chains parallelized over the 10 available CPU cores. The whole MCMC analysis took 8 minutes before converging to a Gellman-Rubin convergence diagnostic of $R-1=0.01$. 
 
 Setting up the plan (see above) and preliminary tasks with \texttt{cmbagent} took another 8-10 minutes and the total cost of the session was \$1.55, for 273843 tokens of which 1803 where completion tokens and the rest prompt tokens (which include all the agent instructions and conversation history). Once the MCMC converged the session was stopped and new sessions where restarted to download the original ACT chains and plot the contours respectively (each costing less than \$0.1). In total the full analysis took 40 minutes, required no human written code, and successfully reproduced the results presented in \cite{ACT:2023kun}. 

The contours overlap almost perfectly, with statistically insignificant differences.\footnote{The remaining difference is caused by slightly different precision settings adopted for the neural network emulators used for the theory code versus precision settings of \texttt{camb} of the original analysis.} Without the MAS, reproducing this analysis ``by hand'' would have taken several hours to even an expert cosmologist. Indeed, one would have had to go through GitHub repositories and papers to collect relevant information, write the configuration files for \texttt{cobaya}, the job scripts for the MCMC and Python code for obtaining the contours. With our MAS, all of this is automated and we are able to reproduce it within a few minutes, without writing a line of code ourselves (all code is written and executed by \texttt{cmbagent}). 

These results are highly encouraging; however, since they are about the same task that we used to develop our system, they are not a demonstration of the usefulness of the system beyond that specific task. To probe its reliability, we apply it to other unrelated tasks.

\begin{figure}
        \centering
        \includegraphics[width=0.8\linewidth]{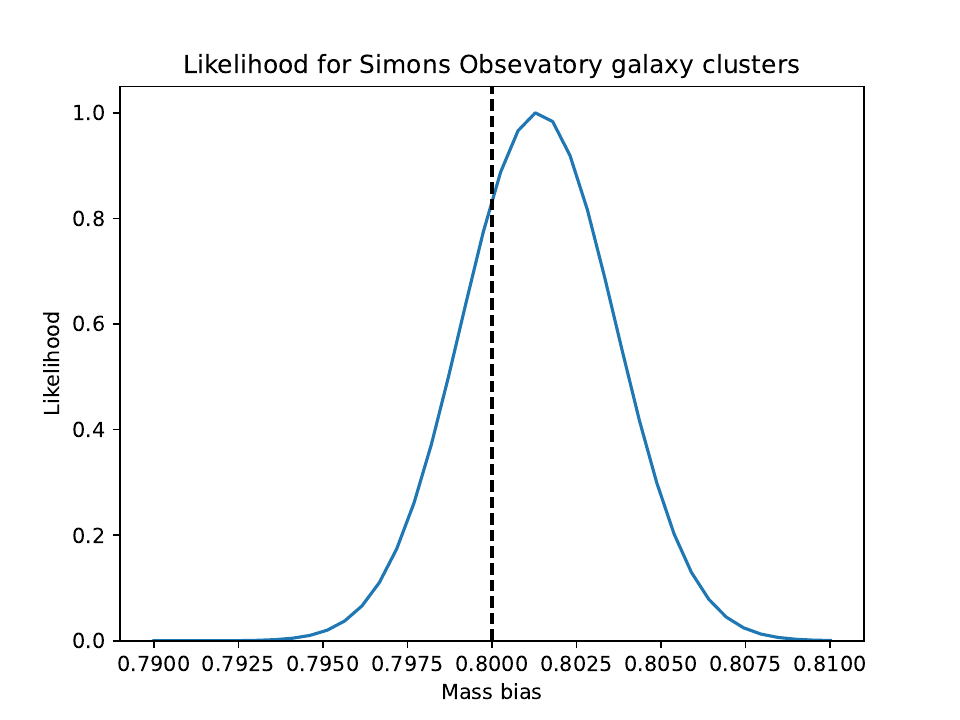}
        \caption{Number-count likelihood for a Simons-Observatory-like simulated catalogue as a function of the mass bias, computed by  \texttt{cosmocnc} via \texttt{cmbagent}. The true input value of the mass bias is shown as the dashed line; it is consistent with the  likelihood constraint obtained by \texttt{cmbagent}. Our GitHub repository contains the \texttt{cosmocnc} agent instructions (see its yaml file).}
        \label{fig:cosmocnc}
\end{figure}

\subsection{Generalization and application to research software} 

\begin{figure}
        \centering
        \includegraphics[width=0.9\linewidth]{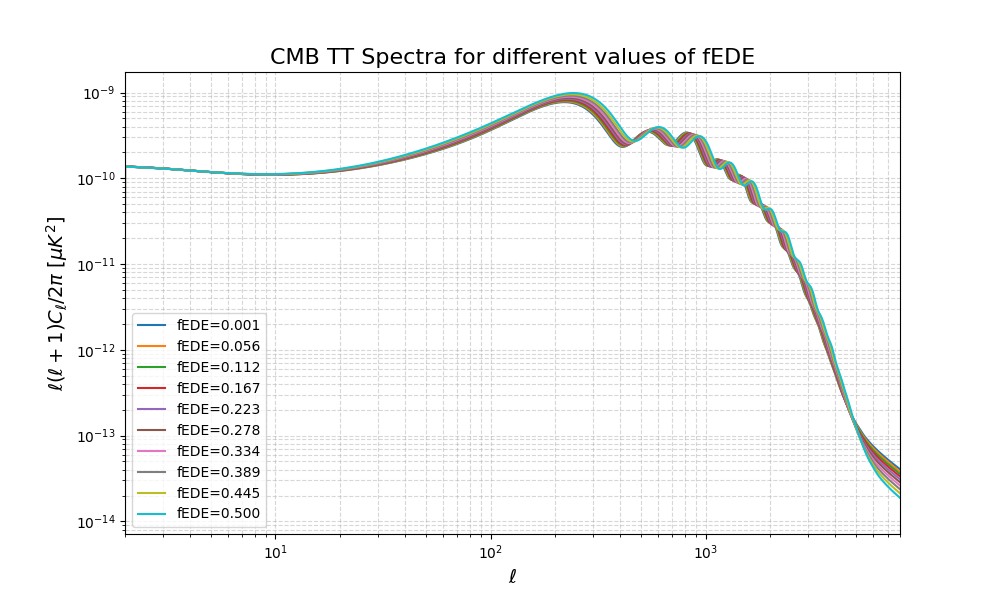}
        \caption{CMB power spectra for ten values of the cosmological parameter $f_\mathrm{EDE}$ computed by \texttt{classy\_sz} via \texttt{cmbagent}. The notebook provided as part of our online documentation contains the full output of the session where Figure \ref{fig:newcode} was made. Our GitHub repository also contains the \texttt{classy\_sz} agent instructions (see yaml file).}
        \label{fig:newcode}
\end{figure}

To demonstrate the usefulness of our example MAS we test it on tasks that have not been considered while developing it.  For example we ask it to compute a cosmological observable for several values of an undocumented parameter $f_{\mathrm{EDE}}$, but implemented in the \texttt{classy\_sz} research software.\footnote{This parameter is part of a modified early universe model known as early dark energy, where $f_{\mathrm{EDE}}$ is a parameter describing the maximum fractional contribution of dark energy to the total cosmic energy budget.} It does it successfully. We show the results on  Fig. \ref{fig:newcode}.  See our online documentation \href{https://cmbagent.readthedocs.io/en/latest/notebooks/cmbagent.html#Compute-models-for-varying-fEDE}{here} for the session where this Figure was obtained.

In addition, we ask it to use the \texttt{cosmocnc} research software to evaluate a  likelihood as a function of one of the key model parameters. Again, it does it successfully, writing Python code to set appropriate values for the different input parameters of \texttt{cosmocnc}, evaluating the likelihood in an efficient way, and plotting the evaluated likelihood. See Fig.\,\ref{fig:cosmocnc} for the results of the calculation and Appendix \ref{app:rs} for the code. 

\section{Limitations and future directions} 

The MAS designed here relies on human feedback at every step outlined by the planner. The user must either give the signal to proceed or to modify the plan, with directions describing which agent to consult for the plan's modification. Future iterations of this code will aim to limit the amount of human input needed, while still maintaining flexibility for the user.  
The package currently assumes the user has done all package and software installation beforehand. A direct extension of our work is to incorporate instructions on how to install cosmology software and packages. Examples of framework for agentic software development that promise to be powerful include \href{https://github.com/All-Hands-AI/OpenHands}{\texttt{OpenHands}} \citep{wang2024openhandsopenplatformai}. This is often a time consuming part and having a MAS to help with it could significantly improve the learning curve for new researchers and help with collaboration on-boarding. 
In this context, other examples of promising LLM-based tools being developed include  \href{https://chatgpt.com/g/g-aYZOjK5zy-chatgaia}{\texttt{ChatGaia}} and \href{https://nolank.ca/astrocoder/}{\texttt{AstroCoder}}.\footnote{See \href{https://nolank.ca}{https://nolank.ca} for further details.}

Nonetheless, it should be noted that, when applied to cosmology, current LLMs frequently produce over-confident, plausible-looking but physically incorrect responses which could easily mislead unexperienced researchers. This is why, at this time, such system may only be useful to experienced researchers that can easily spot when the LLM goes into non-nonsensical directions and stop it in time. Developing agents that can criticize and use logic or reasoning could alleviate these difficulties. One other use could be MAS to serve as a cross-check a human analysis.

An important limitation is token usage and cost. Solving the main task required 15 agent calls, a total of 274483 tokens and costed \$1.55 (see examples \href{https://cmbagent.readthedocs.io/en/latest/notebooks/cmbagent.html#Reproducing-ACT-DR6-lensing-analysis}{here} for details). Less demanding tasks, such as making a plot or designing a plan remain more affordable (we used \$0.1 to produce Figure \ref{fig:contours}, with 6 agent calls). Exploring different models and caching will be key to overcome this limitation.

The RAG part also presents challenges. Retrieval on scientific documentation turns out to be a difficult task when complex tables and equations are involved. Models and pipelines, such as \href{https://docs.llamaindex.ai/en/stable/llama_cloud/llama_parse/}{\texttt{llamaparse}}, are evolving fast. However, even latest versions do not give robust answers. Future work will explore agentic RAG tools for scientific literature, such as \href{https://github.com/Future-House/paper-qa}{\texttt{PaperQA2}} \citep{2024arXiv240913740S}.

Here, we exclusively use GPT-4o. Nonetheless, it is possible to have different agents use different LLMs. Since some LLMs display superior performance in certain domains \citep{chiang2024chatbot}, e.g., for code development, maths, planning, we expect that this will be a promising future direction of our work. A particular class of LLMs that will be of interest are fine-tuned LLMs. Fine-tuning instead of or in addition to RAG could be a promising avenue. Several works in the astronomy community have started to develop fine-tuned LLMs for astrophysics and cosmology; for instance \texttt{cosmosage} \citep{deHaan:2024ybs}, \texttt{AstroLLama} \citep{Nguyen:2023nhp}, or \texttt{AstroMLab} \citep{ting2024astromlab1winsastronomy, 2024arXiv240919750P, dehaan2024astromlab3achievinggpt4o}, and more recently \href{https://github.com/sultan-hassan/CosmoGemma}{\texttt{sultan-hassan/CosmoGemma}}. We plan to study how fine-tuned LLMs behave in our multi-agent systems.

In future work, we plan to explore agentic approaches that use both RAG and fine-tuned LLMs together. 

Another continuation of our research will be on systems with minimal human input, i.e., "Zero-player game", building on recent progress on multi agent reinforcement learning work, see, e.g. \cite{app11114948}, and  \textit{mutual reasoning} \citep{2024arXiv240806195Q}, towards full automation and optimization. The work of \cite{cheng2024trace} is particularly relevant in this context.\footnote{See \href{https://github.com/microsoft/trace}{microsoft/trace} for their code.}  Other directions for optimizing the effectiveness and efficiency of MAS include Monte Carlo Tree Search techniques described in \cite{2024arXiv241008115C}. As described in  \cite{2024arXiv240806292L}, we can even start envisioning autonomous systems making scientific discoveries independently and going all the way through writing scientific publications.
This being said, the greatest  challenge for further automation is that there is no way to confirm outputs without redoing the analysis independently and it is difficult to envision fully automated robust system being deployed for fundamental research in the very near future.

Nonetheless, the most urgent and essential task for such systems to become useful is robust a benchmarking framework. The benchmarking must be done for RAG-type tasks (i.e., Q\&A), for which we plan to build a specific cosmology QA dataset (see for instance  \href{https://huggingface.co/datasets/sultan-hassan/Arxiv_astroph.CO_QA_pairs_2018_2022}{\texttt{astroph.CO\_QA\_pairs}} on HuggingFace) but also for agentic tasks. The latter is still very much an open research question \citep[see, e.g.,][]{2020arXiv200607869P, 2023arXiv231201472B} where reinforcement learning methods for policy optimization play a central role \citep[e.g.,][]{schulman2017, yang2021overviewmultiagentreinforcementlearning, wen2024reinforcinglanguageagentspolicy, marl-book}.

\section*{Acknowledgements}

We are grateful to Ryan Daniels, Adrian Dimitrov, Jo Dunkley, James Fergusson, Simone Ferraro, Sultan Hassan, Nolan Koblischke, Anthony Kremin, Sven Krippendorf, Sigurd Naess,  Frank Qu, Suzanne Staggs, Chi Wang, Ethan Yoo for discussions and/or feedback. We are also grateful to the \href{https://science.ai.cam.ac.uk}{Cambridge Centre for Data-Driven Discovery Accelerate Programme} for support. BDS acknowledges support
from the European Research Council (ERC) under the
European Union’s Horizon 2020 research and innovation
programme (Grant agreement No.~851274) and from an STFC Ernest Rutherford
Fellowship. KS acknowledges support from the National Science Foundation Graduate Research Fellowship Program under Grant No.~DGE 2036197.

\bibliographystyle{apsrev}
\bibliography{refs}

\begin{thebibliography}{43}
\expandafter\ifx\csname natexlab\endcsname\relax\def\natexlab#1{#1}\fi
\expandafter\ifx\csname bibnamefont\endcsname\relax
  \def\bibnamefont#1{#1}\fi
\expandafter\ifx\csname bibfnamefont\endcsname\relax
  \def\bibfnamefont#1{#1}\fi
\expandafter\ifx\csname citenamefont\endcsname\relax
  \def\citenamefont#1{#1}\fi
\expandafter\ifx\csname url\endcsname\relax
  \def\url#1{\texttt{#1}}\fi
\expandafter\ifx\csname urlprefix\endcsname\relax\def\urlprefix{URL }\fi
\providecommand{\bibinfo}[2]{#2}
\providecommand{\eprint}[2][]{\url{#2}}

\bibitem[{\citenamefont{Wu et~al.}(2023)\citenamefont{Wu, Bansal, Zhang, Wu,
  Li, Zhu, Jiang, Zhang, Zhang, Liu et~al.}}]{wu2023autogen}
\bibinfo{author}{\bibfnamefont{Q.}~\bibnamefont{Wu}},
  \bibinfo{author}{\bibfnamefont{G.}~\bibnamefont{Bansal}},
  \bibinfo{author}{\bibfnamefont{J.}~\bibnamefont{Zhang}},
  \bibinfo{author}{\bibfnamefont{Y.}~\bibnamefont{Wu}},
  \bibinfo{author}{\bibfnamefont{B.}~\bibnamefont{Li}},
  \bibinfo{author}{\bibfnamefont{E.}~\bibnamefont{Zhu}},
  \bibinfo{author}{\bibfnamefont{L.}~\bibnamefont{Jiang}},
  \bibinfo{author}{\bibfnamefont{X.}~\bibnamefont{Zhang}},
  \bibinfo{author}{\bibfnamefont{S.}~\bibnamefont{Zhang}},
  \bibinfo{author}{\bibfnamefont{J.}~\bibnamefont{Liu}}, \bibnamefont{et~al.},
  \emph{\bibinfo{title}{Autogen: Enabling next-gen llm applications via
  multi-agent conversation}} (\bibinfo{year}{2023}), \eprint{2308.08155}.

\bibitem[{\citenamefont{{Touvron} et~al.}(2023)\citenamefont{{Touvron},
  {Lavril}, {Izacard}, {Martinet}, {Lachaux}, {Lacroix}, {Rozi{\`e}re},
  {Goyal}, {Hambro}, {Azhar} et~al.}}]{2023arXiv230213971T}
\bibinfo{author}{\bibfnamefont{H.}~\bibnamefont{{Touvron}}},
  \bibinfo{author}{\bibfnamefont{T.}~\bibnamefont{{Lavril}}},
  \bibinfo{author}{\bibfnamefont{G.}~\bibnamefont{{Izacard}}},
  \bibinfo{author}{\bibfnamefont{X.}~\bibnamefont{{Martinet}}},
  \bibinfo{author}{\bibfnamefont{M.-A.} \bibnamefont{{Lachaux}}},
  \bibinfo{author}{\bibfnamefont{T.}~\bibnamefont{{Lacroix}}},
  \bibinfo{author}{\bibfnamefont{B.}~\bibnamefont{{Rozi{\`e}re}}},
  \bibinfo{author}{\bibfnamefont{N.}~\bibnamefont{{Goyal}}},
  \bibinfo{author}{\bibfnamefont{E.}~\bibnamefont{{Hambro}}},
  \bibinfo{author}{\bibfnamefont{F.}~\bibnamefont{{Azhar}}},
  \bibnamefont{et~al.}, \emph{\bibinfo{title}{{LLaMA: Open and Efficient
  Foundation Language Models}}} (\bibinfo{year}{2023}), \eprint{2302.13971}.

\bibitem[{\citenamefont{{Brown} et~al.}(2020)\citenamefont{{Brown}, {Mann},
  {Ryder}, {Subbiah}, {Kaplan}, {Dhariwal}, {Neelakantan}, {Shyam}, {Sastry},
  {Askell} et~al.}}]{2020arXiv200514165B}
\bibinfo{author}{\bibfnamefont{T.~B.} \bibnamefont{{Brown}}},
  \bibinfo{author}{\bibfnamefont{B.}~\bibnamefont{{Mann}}},
  \bibinfo{author}{\bibfnamefont{N.}~\bibnamefont{{Ryder}}},
  \bibinfo{author}{\bibfnamefont{M.}~\bibnamefont{{Subbiah}}},
  \bibinfo{author}{\bibfnamefont{J.}~\bibnamefont{{Kaplan}}},
  \bibinfo{author}{\bibfnamefont{P.}~\bibnamefont{{Dhariwal}}},
  \bibinfo{author}{\bibfnamefont{A.}~\bibnamefont{{Neelakantan}}},
  \bibinfo{author}{\bibfnamefont{P.}~\bibnamefont{{Shyam}}},
  \bibinfo{author}{\bibfnamefont{G.}~\bibnamefont{{Sastry}}},
  \bibinfo{author}{\bibfnamefont{A.}~\bibnamefont{{Askell}}},
  \bibnamefont{et~al.}, \emph{\bibinfo{title}{{Language Models are Few-Shot
  Learners}}} (\bibinfo{year}{2020}), \eprint{2005.14165}.

\bibitem[{\citenamefont{{OpenAI} et~al.}(2023)\citenamefont{{OpenAI}, {Achiam},
  {Adler}, {Agarwal}, {Ahmad}, {Akkaya}, {Leoni Aleman}, {Almeida},
  {Altenschmidt}, {Altman} et~al.}}]{2023arXiv230308774O}
\bibinfo{author}{\bibnamefont{{OpenAI}}},
  \bibinfo{author}{\bibfnamefont{J.}~\bibnamefont{{Achiam}}},
  \bibinfo{author}{\bibfnamefont{S.}~\bibnamefont{{Adler}}},
  \bibinfo{author}{\bibfnamefont{S.}~\bibnamefont{{Agarwal}}},
  \bibinfo{author}{\bibfnamefont{L.}~\bibnamefont{{Ahmad}}},
  \bibinfo{author}{\bibfnamefont{I.}~\bibnamefont{{Akkaya}}},
  \bibinfo{author}{\bibfnamefont{F.}~\bibnamefont{{Leoni Aleman}}},
  \bibinfo{author}{\bibfnamefont{D.}~\bibnamefont{{Almeida}}},
  \bibinfo{author}{\bibfnamefont{J.}~\bibnamefont{{Altenschmidt}}},
  \bibinfo{author}{\bibfnamefont{S.}~\bibnamefont{{Altman}}},
  \bibnamefont{et~al.}, \emph{\bibinfo{title}{{GPT-4 Technical Report}}}
  (\bibinfo{year}{2023}), \eprint{2303.08774}.

\bibitem[{\citenamefont{{Tanoglidis} et~al.}(2023)\citenamefont{{Tanoglidis},
  {Jain}, and {Qu}}}]{2023arXiv231012069T}
\bibinfo{author}{\bibfnamefont{D.}~\bibnamefont{{Tanoglidis}}},
  \bibinfo{author}{\bibfnamefont{B.}~\bibnamefont{{Jain}}}, \bibnamefont{and}
  \bibinfo{author}{\bibfnamefont{H.}~\bibnamefont{{Qu}}},
  \emph{\bibinfo{title}{{Transformers for scientific data: a pedagogical review
  for astronomers}}} (\bibinfo{year}{2023}), \eprint{2310.12069}.

\bibitem[{\citenamefont{{Ghafarollahi} and {Buehler}}(2024)}]{sciagents}
\bibinfo{author}{\bibfnamefont{A.}~\bibnamefont{{Ghafarollahi}}}
  \bibnamefont{and} \bibinfo{author}{\bibfnamefont{M.~J.}
  \bibnamefont{{Buehler}}}, \emph{\bibinfo{title}{{SciAgents: Automating
  scientific discovery through multi-agent intelligent graph reasoning}}}
  (\bibinfo{year}{2024}), \eprint{2409.05556}.

\bibitem[{\citenamefont{Buehler}(2024)}]{buehler2024graphreasoning}
\bibinfo{author}{\bibfnamefont{M.~J.} \bibnamefont{Buehler}},
  \emph{\bibinfo{title}{Accelerating scientific discovery with generative
  knowledge extraction, graph-based representation, and multimodal intelligent
  graph reasoning}} (\bibinfo{year}{2024}),
  \urlprefix\url{http://iopscience.iop.org/article/10.1088/2632-2153/ad7228}.

\bibitem[{\citenamefont{{Sun} et~al.}(2024)\citenamefont{{Sun}, {Ting},
  {Liang}, {Duan}, {Huang}, and {Cai}}}]{mephisto}
\bibinfo{author}{\bibfnamefont{Z.}~\bibnamefont{{Sun}}},
  \bibinfo{author}{\bibfnamefont{Y.-S.} \bibnamefont{{Ting}}},
  \bibinfo{author}{\bibfnamefont{Y.}~\bibnamefont{{Liang}}},
  \bibinfo{author}{\bibfnamefont{N.}~\bibnamefont{{Duan}}},
  \bibinfo{author}{\bibfnamefont{S.}~\bibnamefont{{Huang}}}, \bibnamefont{and}
  \bibinfo{author}{\bibfnamefont{Z.}~\bibnamefont{{Cai}}},
  \emph{\bibinfo{title}{{Interpreting Multi-band Galaxy Observations with Large
  Language Model-Based Agents}}} (\bibinfo{year}{2024}), \eprint{2409.14807}.

\bibitem[{\citenamefont{Madhavacheril et~al.}(2024)}]{ACT:2023kun}
\bibinfo{author}{\bibfnamefont{M.~S.} \bibnamefont{Madhavacheril}}
  \bibnamefont{et~al.} (\bibinfo{collaboration}{ACT}),
  \emph{\bibinfo{title}{{The Atacama Cosmology Telescope: DR6 Gravitational
  Lensing Map and Cosmological Parameters}}} (\bibinfo{year}{2024}),
  \eprint{2304.05203}.

\bibitem[{\citenamefont{Qu et~al.}(2024)}]{ACT:2023dou}
\bibinfo{author}{\bibfnamefont{F.~J.} \bibnamefont{Qu}} \bibnamefont{et~al.}
  (\bibinfo{collaboration}{ACT}), \emph{\bibinfo{title}{{The Atacama Cosmology
  Telescope: A Measurement of the DR6 CMB Lensing Power Spectrum and Its
  Implications for Structure Growth}}} (\bibinfo{year}{2024}),
  \eprint{2304.05202}.

\bibitem[{\citenamefont{MacCrann et~al.}(2023)\citenamefont{MacCrann, Sherwin,
  Qu, Namikawa, Madhavacheril, Abril-Cabezas, An, Austermann, Battaglia,
  Battistelli et~al.}}]{maccrann2023atacamacosmologytelescopemitigating}
\bibinfo{author}{\bibfnamefont{N.}~\bibnamefont{MacCrann}},
  \bibinfo{author}{\bibfnamefont{B.~D.} \bibnamefont{Sherwin}},
  \bibinfo{author}{\bibfnamefont{F.~J.} \bibnamefont{Qu}},
  \bibinfo{author}{\bibfnamefont{T.}~\bibnamefont{Namikawa}},
  \bibinfo{author}{\bibfnamefont{M.~S.} \bibnamefont{Madhavacheril}},
  \bibinfo{author}{\bibfnamefont{I.}~\bibnamefont{Abril-Cabezas}},
  \bibinfo{author}{\bibfnamefont{R.}~\bibnamefont{An}},
  \bibinfo{author}{\bibfnamefont{J.~E.} \bibnamefont{Austermann}},
  \bibinfo{author}{\bibfnamefont{N.}~\bibnamefont{Battaglia}},
  \bibinfo{author}{\bibfnamefont{E.~S.} \bibnamefont{Battistelli}},
  \bibnamefont{et~al.}, \emph{\bibinfo{title}{The atacama cosmology telescope:
  Mitigating the impact of extragalactic foregrounds for the dr6 cmb lensing
  analysis}} (\bibinfo{year}{2023}), \eprint{2304.05196},
  \urlprefix\url{https://arxiv.org/abs/2304.05196}.

\bibitem[{\citenamefont{Lewis et~al.}(2000)\citenamefont{Lewis, Challinor, and
  Lasenby}}]{Lewis:1999bs}
\bibinfo{author}{\bibfnamefont{A.}~\bibnamefont{Lewis}},
  \bibinfo{author}{\bibfnamefont{A.}~\bibnamefont{Challinor}},
  \bibnamefont{and} \bibinfo{author}{\bibfnamefont{A.}~\bibnamefont{Lasenby}},
  \emph{\bibinfo{title}{{Efficient computation of CMB anisotropies in closed
  FRW models}}} (\bibinfo{year}{2000}), \eprint{astro-ph/9911177}.

\bibitem[{\citenamefont{Blas et~al.}(2011)\citenamefont{Blas, Lesgourgues, and
  Tram}}]{class}
\bibinfo{author}{\bibfnamefont{D.}~\bibnamefont{Blas}},
  \bibinfo{author}{\bibfnamefont{J.}~\bibnamefont{Lesgourgues}},
  \bibnamefont{and} \bibinfo{author}{\bibfnamefont{T.}~\bibnamefont{Tram}},
  \emph{\bibinfo{title}{The cosmic linear anisotropy solving system (class).
  part ii: Approximation schemes}} (\bibinfo{year}{2011}), \eprint{1104.2933}.

\bibitem[{\citenamefont{Torrado and Lewis}(2021)}]{Torrado:2020dgo}
\bibinfo{author}{\bibfnamefont{J.}~\bibnamefont{Torrado}} \bibnamefont{and}
  \bibinfo{author}{\bibfnamefont{A.}~\bibnamefont{Lewis}},
  \emph{\bibinfo{title}{{Cobaya: Code for Bayesian Analysis of hierarchical
  physical models}}} (\bibinfo{year}{2021}), \eprint{2005.05290}.

\bibitem[{\citenamefont{Lewis}(2019)}]{Lewis:2019xzd}
\bibinfo{author}{\bibfnamefont{A.}~\bibnamefont{Lewis}},
  \emph{\bibinfo{title}{{GetDist: a Python package for analysing Monte Carlo
  samples}}} (\bibinfo{year}{2019}), \eprint{1910.13970}.

\bibitem[{\citenamefont{Bolliet et~al.}(2024)}]{Bolliet:2023eob}
\bibinfo{author}{\bibfnamefont{B.}~\bibnamefont{Bolliet}} \bibnamefont{et~al.},
  \emph{\bibinfo{title}{{class\_sz I: Overview}}} (\bibinfo{year}{2024}),
  \eprint{2310.18482}.

\bibitem[{\citenamefont{Zubeldia et~al.}(2024)\citenamefont{Zubeldia, Melin,
  Chluba, and Battye}}]{Zubeldia:2024zow}
\bibinfo{author}{\bibfnamefont{I.~n.} \bibnamefont{Zubeldia}},
  \bibinfo{author}{\bibfnamefont{J.-B.} \bibnamefont{Melin}},
  \bibinfo{author}{\bibfnamefont{J.}~\bibnamefont{Chluba}}, \bibnamefont{and}
  \bibinfo{author}{\bibfnamefont{R.}~\bibnamefont{Battye}},
  \emph{\bibinfo{title}{{The Planck SZiFi catalogues: a new set of Planck
  catalogues of Sunyaev-Zeldovich-detected galaxy clusters}}}
  (\bibinfo{year}{2024}), \eprint{2408.06189}.

\bibitem[{\citenamefont{{Abadi} et~al.}(2016)\citenamefont{{Abadi}, {Barham},
  {Chen}, {Chen}, {Davis}, {Dean}, {Devin}, {Ghemawat}, {Irving}, {Isard}
  et~al.}}]{2016arXiv160508695A}
\bibinfo{author}{\bibfnamefont{M.}~\bibnamefont{{Abadi}}},
  \bibinfo{author}{\bibfnamefont{P.}~\bibnamefont{{Barham}}},
  \bibinfo{author}{\bibfnamefont{J.}~\bibnamefont{{Chen}}},
  \bibinfo{author}{\bibfnamefont{Z.}~\bibnamefont{{Chen}}},
  \bibinfo{author}{\bibfnamefont{A.}~\bibnamefont{{Davis}}},
  \bibinfo{author}{\bibfnamefont{J.}~\bibnamefont{{Dean}}},
  \bibinfo{author}{\bibfnamefont{M.}~\bibnamefont{{Devin}}},
  \bibinfo{author}{\bibfnamefont{S.}~\bibnamefont{{Ghemawat}}},
  \bibinfo{author}{\bibfnamefont{G.}~\bibnamefont{{Irving}}},
  \bibinfo{author}{\bibfnamefont{M.}~\bibnamefont{{Isard}}},
  \bibnamefont{et~al.}, \emph{\bibinfo{title}{{TensorFlow: A system for
  large-scale machine learning}}} (\bibinfo{year}{2016}), \eprint{1605.08695}.

\bibitem[{\citenamefont{Spurio-Mancini
  et~al.}(2022)\citenamefont{Spurio-Mancini, Piras, Alsing, Joachimi, and
  Hobson}}]{Spurio_Mancini_2022}
\bibinfo{author}{\bibfnamefont{A.}~\bibnamefont{Spurio-Mancini}},
  \bibinfo{author}{\bibfnamefont{D.}~\bibnamefont{Piras}},
  \bibinfo{author}{\bibfnamefont{J.}~\bibnamefont{Alsing}},
  \bibinfo{author}{\bibfnamefont{B.}~\bibnamefont{Joachimi}}, \bibnamefont{and}
  \bibinfo{author}{\bibfnamefont{M.~P.} \bibnamefont{Hobson}},
  \emph{\bibinfo{title}{<scp>cosmopower</scp>: emulating cosmological power
  spectra for accelerated bayesian inference from next-generation surveys}}
  (\bibinfo{year}{2022}), \eprint{2106.03846}.

\bibitem[{\citenamefont{{Bolliet} et~al.}(2024)\citenamefont{{Bolliet}, {Spurio
  Mancini}, {Hill}, {Madhavacheril}, {Jense}, {Calabrese}, and
  {Dunkley}}}]{Bolliet:2023sst}
\bibinfo{author}{\bibfnamefont{B.}~\bibnamefont{{Bolliet}}},
  \bibinfo{author}{\bibfnamefont{A.}~\bibnamefont{{Spurio Mancini}}},
  \bibinfo{author}{\bibfnamefont{J.~C.} \bibnamefont{{Hill}}},
  \bibinfo{author}{\bibfnamefont{M.}~\bibnamefont{{Madhavacheril}}},
  \bibinfo{author}{\bibfnamefont{H.~T.} \bibnamefont{{Jense}}},
  \bibinfo{author}{\bibfnamefont{E.}~\bibnamefont{{Calabrese}}},
  \bibnamefont{and}
  \bibinfo{author}{\bibfnamefont{J.}~\bibnamefont{{Dunkley}}},
  \emph{\bibinfo{title}{{High-accuracy emulators for observables in
  {\ensuremath{\Lambda}}CDM, N$_{eff}$,
  {\ensuremath{\Sigma}}m$_{{\ensuremath{\nu}}}$, and w cosmologies}}}
  (\bibinfo{year}{2024}), \eprint{2303.01591}.

\bibitem[{\citenamefont{{Lewis} et~al.}(2020)\citenamefont{{Lewis}, {Perez},
  {Piktus}, {Petroni}, {Karpukhin}, {Goyal}, {K{\"u}ttler}, {Lewis}, {Yih},
  {Rockt{\"a}schel} et~al.}}]{RAG}
\bibinfo{author}{\bibfnamefont{P.}~\bibnamefont{{Lewis}}},
  \bibinfo{author}{\bibfnamefont{E.}~\bibnamefont{{Perez}}},
  \bibinfo{author}{\bibfnamefont{A.}~\bibnamefont{{Piktus}}},
  \bibinfo{author}{\bibfnamefont{F.}~\bibnamefont{{Petroni}}},
  \bibinfo{author}{\bibfnamefont{V.}~\bibnamefont{{Karpukhin}}},
  \bibinfo{author}{\bibfnamefont{N.}~\bibnamefont{{Goyal}}},
  \bibinfo{author}{\bibfnamefont{H.}~\bibnamefont{{K{\"u}ttler}}},
  \bibinfo{author}{\bibfnamefont{M.}~\bibnamefont{{Lewis}}},
  \bibinfo{author}{\bibfnamefont{W.-t.} \bibnamefont{{Yih}}},
  \bibinfo{author}{\bibfnamefont{T.}~\bibnamefont{{Rockt{\"a}schel}}},
  \bibnamefont{et~al.}, \emph{\bibinfo{title}{{Retrieval-Augmented Generation
  for Knowledge-Intensive NLP Tasks}}} (\bibinfo{year}{2020}),
  \eprint{2005.11401}.

\bibitem[{\citenamefont{Aghanim et~al.}(2020)}]{Planck:2018vyg}
\bibinfo{author}{\bibfnamefont{N.}~\bibnamefont{Aghanim}} \bibnamefont{et~al.}
  (\bibinfo{collaboration}{Planck}), \emph{\bibinfo{title}{{Planck 2018
  results. VI. Cosmological parameters}}} (\bibinfo{year}{2020}),
  \bibinfo{note}{[Erratum: Astron.Astrophys. 652, C4 (2021)]},
  \eprint{1807.06209}.

\bibitem[{\citenamefont{Carron et~al.}(2022)\citenamefont{Carron, Mirmelstein,
  and Lewis}}]{Carron:2022eyg}
\bibinfo{author}{\bibfnamefont{J.}~\bibnamefont{Carron}},
  \bibinfo{author}{\bibfnamefont{M.}~\bibnamefont{Mirmelstein}},
  \bibnamefont{and} \bibinfo{author}{\bibfnamefont{A.}~\bibnamefont{Lewis}},
  \emph{\bibinfo{title}{{CMB lensing from Planck PR4~maps}}}
  (\bibinfo{year}{2022}), \eprint{2206.07773}.

\bibitem[{\citenamefont{Tristram et~al.}(2024)}]{Tristram:2023haj}
\bibinfo{author}{\bibfnamefont{M.}~\bibnamefont{Tristram}}
  \bibnamefont{et~al.}, \emph{\bibinfo{title}{{Cosmological parameters derived
  from the final Planck data release (PR4)}}} (\bibinfo{year}{2024}),
  \eprint{2309.10034}.

\bibitem[{\citenamefont{Wang et~al.}(2024)\citenamefont{Wang, Li, Song, Xu,
  Tang, Zhuge, Pan, Song, Li, Singh et~al.}}]{wang2024openhandsopenplatformai}
\bibinfo{author}{\bibfnamefont{X.}~\bibnamefont{Wang}},
  \bibinfo{author}{\bibfnamefont{B.}~\bibnamefont{Li}},
  \bibinfo{author}{\bibfnamefont{Y.}~\bibnamefont{Song}},
  \bibinfo{author}{\bibfnamefont{F.~F.} \bibnamefont{Xu}},
  \bibinfo{author}{\bibfnamefont{X.}~\bibnamefont{Tang}},
  \bibinfo{author}{\bibfnamefont{M.}~\bibnamefont{Zhuge}},
  \bibinfo{author}{\bibfnamefont{J.}~\bibnamefont{Pan}},
  \bibinfo{author}{\bibfnamefont{Y.}~\bibnamefont{Song}},
  \bibinfo{author}{\bibfnamefont{B.}~\bibnamefont{Li}},
  \bibinfo{author}{\bibfnamefont{J.}~\bibnamefont{Singh}},
  \bibnamefont{et~al.}, \emph{\bibinfo{title}{Openhands: An open platform for
  ai software developers as generalist agents}} (\bibinfo{year}{2024}),
  \eprint{2407.16741}.

\bibitem[{\citenamefont{{Skarlinski} et~al.}(2024)\citenamefont{{Skarlinski},
  {Cox}, {Laurent}, {Braza}, {Hinks}, {Hammerling}, {Ponnapati}, {Rodriques},
  and {White}}}]{2024arXiv240913740S}
\bibinfo{author}{\bibfnamefont{M.~D.} \bibnamefont{{Skarlinski}}},
  \bibinfo{author}{\bibfnamefont{S.}~\bibnamefont{{Cox}}},
  \bibinfo{author}{\bibfnamefont{J.~M.} \bibnamefont{{Laurent}}},
  \bibinfo{author}{\bibfnamefont{J.~D.} \bibnamefont{{Braza}}},
  \bibinfo{author}{\bibfnamefont{M.}~\bibnamefont{{Hinks}}},
  \bibinfo{author}{\bibfnamefont{M.~J.} \bibnamefont{{Hammerling}}},
  \bibinfo{author}{\bibfnamefont{M.}~\bibnamefont{{Ponnapati}}},
  \bibinfo{author}{\bibfnamefont{S.~G.} \bibnamefont{{Rodriques}}},
  \bibnamefont{and} \bibinfo{author}{\bibfnamefont{A.~D.}
  \bibnamefont{{White}}}, \emph{\bibinfo{title}{{Language agents achieve
  superhuman synthesis of scientific knowledge}}} (\bibinfo{year}{2024}),
  \eprint{2409.13740}.

\bibitem[{\citenamefont{Chiang et~al.}(2024)\citenamefont{Chiang, Zheng, Sheng,
  Angelopoulos, Li, Li, Zhang, Zhu, Jordan, Gonzalez
  et~al.}}]{chiang2024chatbot}
\bibinfo{author}{\bibfnamefont{W.-L.} \bibnamefont{Chiang}},
  \bibinfo{author}{\bibfnamefont{L.}~\bibnamefont{Zheng}},
  \bibinfo{author}{\bibfnamefont{Y.}~\bibnamefont{Sheng}},
  \bibinfo{author}{\bibfnamefont{A.~N.} \bibnamefont{Angelopoulos}},
  \bibinfo{author}{\bibfnamefont{T.}~\bibnamefont{Li}},
  \bibinfo{author}{\bibfnamefont{D.}~\bibnamefont{Li}},
  \bibinfo{author}{\bibfnamefont{H.}~\bibnamefont{Zhang}},
  \bibinfo{author}{\bibfnamefont{B.}~\bibnamefont{Zhu}},
  \bibinfo{author}{\bibfnamefont{M.}~\bibnamefont{Jordan}},
  \bibinfo{author}{\bibfnamefont{J.~E.} \bibnamefont{Gonzalez}},
  \bibnamefont{et~al.}, \emph{\bibinfo{title}{Chatbot arena: An open platform
  for evaluating llms by human preference}} (\bibinfo{year}{2024}),
  \eprint{2403.04132}.

\bibitem[{\citenamefont{de~Haan}(2024)}]{deHaan:2024ybs}
\bibinfo{author}{\bibfnamefont{T.}~\bibnamefont{de~Haan}},
  \emph{\bibinfo{title}{{cosmosage: A Natural-Language Assistant for
  Cosmologists}}} (\bibinfo{year}{2024}), \eprint{2407.04420}.

\bibitem[{\citenamefont{Nguyen et~al.}(2023)}]{Nguyen:2023nhp}
\bibinfo{author}{\bibfnamefont{T.~D.} \bibnamefont{Nguyen}}
  \bibnamefont{et~al.}, \emph{\bibinfo{title}{{AstroLLaMA: Towards Specialized
  Foundation Models in Astronomy}}} (\bibinfo{year}{2023}),
  \eprint{2309.06126}.

\bibitem[{\citenamefont{Ting et~al.}(2024)\citenamefont{Ting, Nguyen, Ghosal,
  Pan, Arora, Sun, de~Haan, Ramachandra, Wells, Madireddy
  et~al.}}]{ting2024astromlab1winsastronomy}
\bibinfo{author}{\bibfnamefont{Y.-S.} \bibnamefont{Ting}},
  \bibinfo{author}{\bibfnamefont{T.~D.} \bibnamefont{Nguyen}},
  \bibinfo{author}{\bibfnamefont{T.}~\bibnamefont{Ghosal}},
  \bibinfo{author}{\bibfnamefont{R.}~\bibnamefont{Pan}},
  \bibinfo{author}{\bibfnamefont{H.}~\bibnamefont{Arora}},
  \bibinfo{author}{\bibfnamefont{Z.}~\bibnamefont{Sun}},
  \bibinfo{author}{\bibfnamefont{T.}~\bibnamefont{de~Haan}},
  \bibinfo{author}{\bibfnamefont{N.}~\bibnamefont{Ramachandra}},
  \bibinfo{author}{\bibfnamefont{A.}~\bibnamefont{Wells}},
  \bibinfo{author}{\bibfnamefont{S.}~\bibnamefont{Madireddy}},
  \bibnamefont{et~al.}, \emph{\bibinfo{title}{Astromlab 1: Who wins astronomy
  jeopardy!?}} (\bibinfo{year}{2024}), \eprint{2407.11194}.

\bibitem[{\citenamefont{{Pan} et~al.}(2024)\citenamefont{{Pan}, {Dung Nguyen},
  {Arora}, {Accomazzi}, {Ghosal}, and {Ting}}}]{2024arXiv240919750P}
\bibinfo{author}{\bibfnamefont{R.}~\bibnamefont{{Pan}}},
  \bibinfo{author}{\bibfnamefont{T.}~\bibnamefont{{Dung Nguyen}}},
  \bibinfo{author}{\bibfnamefont{H.}~\bibnamefont{{Arora}}},
  \bibinfo{author}{\bibfnamefont{A.}~\bibnamefont{{Accomazzi}}},
  \bibinfo{author}{\bibfnamefont{T.}~\bibnamefont{{Ghosal}}}, \bibnamefont{and}
  \bibinfo{author}{\bibfnamefont{Y.-S.} \bibnamefont{{Ting}}},
  \emph{\bibinfo{title}{{AstroMLab 2: AstroLLaMA-2-70B Model and Benchmarking
  Specialised LLMs for Astronomy}}} (\bibinfo{year}{2024}),
  \eprint{2409.19750}.

\bibitem[{\citenamefont{de~Haan et~al.}(2024)\citenamefont{de~Haan, Ting,
  Ghosal, Nguyen, Accomazzi, Wells, Ramachandra, Pan, and
  Sun}}]{dehaan2024astromlab3achievinggpt4o}
\bibinfo{author}{\bibfnamefont{T.}~\bibnamefont{de~Haan}},
  \bibinfo{author}{\bibfnamefont{Y.-S.} \bibnamefont{Ting}},
  \bibinfo{author}{\bibfnamefont{T.}~\bibnamefont{Ghosal}},
  \bibinfo{author}{\bibfnamefont{T.~D.} \bibnamefont{Nguyen}},
  \bibinfo{author}{\bibfnamefont{A.}~\bibnamefont{Accomazzi}},
  \bibinfo{author}{\bibfnamefont{A.}~\bibnamefont{Wells}},
  \bibinfo{author}{\bibfnamefont{N.}~\bibnamefont{Ramachandra}},
  \bibinfo{author}{\bibfnamefont{R.}~\bibnamefont{Pan}}, \bibnamefont{and}
  \bibinfo{author}{\bibfnamefont{Z.}~\bibnamefont{Sun}},
  \emph{\bibinfo{title}{Astromlab 3: Achieving gpt-4o level performance in
  astronomy with a specialized 8b-parameter large language model}}
  (\bibinfo{year}{2024}), \eprint{2411.09012}.

\bibitem[{\citenamefont{Canese et~al.}(2021)\citenamefont{Canese, Cardarilli,
  Di~Nunzio, Fazzolari, Giardino, Re, and Spanò}}]{app11114948}
\bibinfo{author}{\bibfnamefont{L.}~\bibnamefont{Canese}},
  \bibinfo{author}{\bibfnamefont{G.~C.} \bibnamefont{Cardarilli}},
  \bibinfo{author}{\bibfnamefont{L.}~\bibnamefont{Di~Nunzio}},
  \bibinfo{author}{\bibfnamefont{R.}~\bibnamefont{Fazzolari}},
  \bibinfo{author}{\bibfnamefont{D.}~\bibnamefont{Giardino}},
  \bibinfo{author}{\bibfnamefont{M.}~\bibnamefont{Re}}, \bibnamefont{and}
  \bibinfo{author}{\bibfnamefont{S.}~\bibnamefont{Spanò}},
  \emph{\bibinfo{title}{Multi-agent reinforcement learning: A review of
  challenges and applications}} (\bibinfo{year}{2021}),
  \urlprefix\url{https://www.mdpi.com/2076-3417/11/11/4948}.

\bibitem[{\citenamefont{{Qi} et~al.}(2024)\citenamefont{{Qi}, {Ma}, {Xu}, {Lyna
  Zhang}, {Yang}, and {Yang}}}]{2024arXiv240806195Q}
\bibinfo{author}{\bibfnamefont{Z.}~\bibnamefont{{Qi}}},
  \bibinfo{author}{\bibfnamefont{M.}~\bibnamefont{{Ma}}},
  \bibinfo{author}{\bibfnamefont{J.}~\bibnamefont{{Xu}}},
  \bibinfo{author}{\bibfnamefont{L.}~\bibnamefont{{Lyna Zhang}}},
  \bibinfo{author}{\bibfnamefont{F.}~\bibnamefont{{Yang}}}, \bibnamefont{and}
  \bibinfo{author}{\bibfnamefont{M.}~\bibnamefont{{Yang}}},
  \emph{\bibinfo{title}{{Mutual Reasoning Makes Smaller LLMs Stronger
  Problem-Solvers}}} (\bibinfo{year}{2024}), \eprint{2408.06195}.

\bibitem[{\citenamefont{Cheng et~al.}(2024)\citenamefont{Cheng, Nie, and
  Swaminathan}}]{cheng2024trace}
\bibinfo{author}{\bibfnamefont{C.-A.} \bibnamefont{Cheng}},
  \bibinfo{author}{\bibfnamefont{A.}~\bibnamefont{Nie}}, \bibnamefont{and}
  \bibinfo{author}{\bibfnamefont{A.}~\bibnamefont{Swaminathan}},
  \emph{\bibinfo{title}{Trace is the next autodiff: Generative optimization
  with rich feedback, execution traces, and llms}} (\bibinfo{year}{2024}),
  \eprint{2406.16218}.

\bibitem[{\citenamefont{{Chen} et~al.}(2024)\citenamefont{{Chen}, {Yuan},
  {Qian}, {Yang}, {Liu}, and {Sun}}}]{2024arXiv241008115C}
\bibinfo{author}{\bibfnamefont{W.}~\bibnamefont{{Chen}}},
  \bibinfo{author}{\bibfnamefont{J.}~\bibnamefont{{Yuan}}},
  \bibinfo{author}{\bibfnamefont{C.}~\bibnamefont{{Qian}}},
  \bibinfo{author}{\bibfnamefont{C.}~\bibnamefont{{Yang}}},
  \bibinfo{author}{\bibfnamefont{Z.}~\bibnamefont{{Liu}}}, \bibnamefont{and}
  \bibinfo{author}{\bibfnamefont{M.}~\bibnamefont{{Sun}}},
  \emph{\bibinfo{title}{Optima: Optimizing effectiveness and efficiency for
  llm-based multi-agent system}} (\bibinfo{year}{2024}), \eprint{2410.08115}.

\bibitem[{\citenamefont{{Lu} et~al.}(2024)\citenamefont{{Lu}, {Lu}, {Tjarko
  Lange}, {Foerster}, {Clune}, and {Ha}}}]{2024arXiv240806292L}
\bibinfo{author}{\bibfnamefont{C.}~\bibnamefont{{Lu}}},
  \bibinfo{author}{\bibfnamefont{C.}~\bibnamefont{{Lu}}},
  \bibinfo{author}{\bibfnamefont{R.}~\bibnamefont{{Tjarko Lange}}},
  \bibinfo{author}{\bibfnamefont{J.}~\bibnamefont{{Foerster}}},
  \bibinfo{author}{\bibfnamefont{J.}~\bibnamefont{{Clune}}}, \bibnamefont{and}
  \bibinfo{author}{\bibfnamefont{D.}~\bibnamefont{{Ha}}},
  \emph{\bibinfo{title}{{The AI Scientist: Towards Fully Automated Open-Ended
  Scientific Discovery}}} (\bibinfo{year}{2024}), \eprint{2408.06292}.

\bibitem[{\citenamefont{{Papoudakis} et~al.}(2020)\citenamefont{{Papoudakis},
  {Christianos}, {Sch{\"a}fer}, and {Albrecht}}}]{2020arXiv200607869P}
\bibinfo{author}{\bibfnamefont{G.}~\bibnamefont{{Papoudakis}}},
  \bibinfo{author}{\bibfnamefont{F.}~\bibnamefont{{Christianos}}},
  \bibinfo{author}{\bibfnamefont{L.}~\bibnamefont{{Sch{\"a}fer}}},
  \bibnamefont{and} \bibinfo{author}{\bibfnamefont{S.~V.}
  \bibnamefont{{Albrecht}}}, \emph{\bibinfo{title}{{Benchmarking Multi-Agent
  Deep Reinforcement Learning Algorithms in Cooperative Tasks}}}
  (\bibinfo{year}{2020}), \eprint{2006.07869}.

\bibitem[{\citenamefont{{Bettini} et~al.}(2023)\citenamefont{{Bettini},
  {Prorok}, and {Moens}}}]{2023arXiv231201472B}
\bibinfo{author}{\bibfnamefont{M.}~\bibnamefont{{Bettini}}},
  \bibinfo{author}{\bibfnamefont{A.}~\bibnamefont{{Prorok}}}, \bibnamefont{and}
  \bibinfo{author}{\bibfnamefont{V.}~\bibnamefont{{Moens}}},
  \emph{\bibinfo{title}{{BenchMARL: Benchmarking Multi-Agent Reinforcement
  Learning}}} (\bibinfo{year}{2023}), \eprint{2312.01472}.

\bibitem[{\citenamefont{Schulman et~al.}(2017)\citenamefont{Schulman, Wolski,
  Dhariwal, Radford, and Klimov}}]{schulman2017}
\bibinfo{author}{\bibfnamefont{J.}~\bibnamefont{Schulman}},
  \bibinfo{author}{\bibfnamefont{F.}~\bibnamefont{Wolski}},
  \bibinfo{author}{\bibfnamefont{P.}~\bibnamefont{Dhariwal}},
  \bibinfo{author}{\bibfnamefont{A.}~\bibnamefont{Radford}}, \bibnamefont{and}
  \bibinfo{author}{\bibfnamefont{O.}~\bibnamefont{Klimov}},
  \emph{\bibinfo{title}{Proximal policy optimization algorithms}}
  (\bibinfo{year}{2017}), \eprint{1707.06347},
  \urlprefix\url{https://arxiv.org/abs/1707.06347}.

\bibitem[{\citenamefont{Yang and
  Wang}(2021)}]{yang2021overviewmultiagentreinforcementlearning}
\bibinfo{author}{\bibfnamefont{Y.}~\bibnamefont{Yang}} \bibnamefont{and}
  \bibinfo{author}{\bibfnamefont{J.}~\bibnamefont{Wang}},
  \emph{\bibinfo{title}{An overview of multi-agent reinforcement learning from
  game theoretical perspective}} (\bibinfo{year}{2021}), \eprint{2011.00583},
  \urlprefix\url{https://arxiv.org/abs/2011.00583}.

\bibitem[{\citenamefont{Wen et~al.}(2024)\citenamefont{Wen, Wan, Zhang, Wang,
  and Wen}}]{wen2024reinforcinglanguageagentspolicy}
\bibinfo{author}{\bibfnamefont{M.}~\bibnamefont{Wen}},
  \bibinfo{author}{\bibfnamefont{Z.}~\bibnamefont{Wan}},
  \bibinfo{author}{\bibfnamefont{W.}~\bibnamefont{Zhang}},
  \bibinfo{author}{\bibfnamefont{J.}~\bibnamefont{Wang}}, \bibnamefont{and}
  \bibinfo{author}{\bibfnamefont{Y.}~\bibnamefont{Wen}},
  \emph{\bibinfo{title}{Reinforcing language agents via policy optimization
  with action decomposition}} (\bibinfo{year}{2024}), \eprint{2405.15821},
  \urlprefix\url{https://arxiv.org/abs/2405.15821}.

\bibitem[{\citenamefont{Albrecht et~al.}(2024)\citenamefont{Albrecht,
  Christianos, and Sch\"afer}}]{marl-book}
\bibinfo{author}{\bibfnamefont{S.~V.} \bibnamefont{Albrecht}},
  \bibinfo{author}{\bibfnamefont{F.}~\bibnamefont{Christianos}},
  \bibnamefont{and}
  \bibinfo{author}{\bibfnamefont{L.}~\bibnamefont{Sch\"afer}},
  \emph{\bibinfo{title}{Multi-Agent Reinforcement Learning: Foundations and
  Modern Approaches}} (\bibinfo{publisher}{MIT Press}, \bibinfo{year}{2024}),
  \urlprefix\url{https://www.marl-book.com}.

\end{thebibliography}

\appendix
\newpage

\section{Research software example: cosmocnc}\label{app:rs}

 The code boxes below show how the \texttt{cosmocnc} session to make Figure \ref{fig:cosmocnc} was started and the code that \texttt{cmbagent} provided as a result.

\definecolor{mygreen}{rgb}{0,0.6,0}
\definecolor{mygray}{rgb}{0.5,0.5,0.5}
\definecolor{mymauve}{rgb}{0.58,0,0.82}

\lstset{ %
  backgroundcolor=\color{white},   
  basicstyle=\footnotesize\ttfamily, 
  breaklines=true,                 
  captionpos=b,                    
  commentstyle=\color{mygreen},    
  keywordstyle=\color{blue},       
  stringstyle=\color{mymauve},     
  numbers=left,                    
  numberstyle=\tiny\color{mygray}, 
  stepnumber=1,                    
  numbersep=10pt,                  
  tabsize=4,                       
  showspaces=false,                
  showstringspaces=false,          
  showtabs=false,                  
  frame=single,                    
  rulecolor=\color{black},         
  title=\lstname                   
}

\begin{lstlisting}[language=Python, caption=Python code showing the start of session where Fig 4 was produced with the resulting code shown hereafter.]
from cmbagent import CMBAgent

agent_temperature = {
    'cosmocnc_agent': 0.000001
}

agent_top_p = {
    'cosmocnc_agent': 0.1,
}

cmbagent = CMBAgent(agent_list = ['cosmocnc'],
                    verbose=True,
                    agent_instructions = {},
                    agent_temperature = agent_temperature,
                    agent_top_p = agent_top_p
                   )

task = """Use cosmocnc to write code to compute the unbinned log-likelihood 
for the "SO_sim_0" catalogue, only for one mass observable, "q_so_sim", 
for 40 values of "bias_sz", linearly spaced between 0.79 and 0.81. 
The code must plot the exponential of the log-likelihood, normalising it 
to one at its highest value, and save the plot as a pdf file and the code 
as a .py file.
Plan:
1. Set all the cosmocnc parameters explicitly in the code, to the appropriate values.
2. Save the code in output/results/ as a .py file.
 """

cmbagent.solve(task)
\end{lstlisting}

\begin{lstlisting}[language=Python, caption=Python code produced by cmbagent  which makes Figure 4 and obtained as a result of session started in Listing 1 above.]
import numpy as np
import matplotlib.pyplot as plt
import cosmocnc

# Set cosmocnc parameters
cnc_params = cosmocnc.cnc_params_default
scal_rel_params = cosmocnc.scaling_relation_params_default
cosmo_params = cosmocnc.cosmo_params_default

# Catalogue and observables
cnc_params["cluster_catalogue"] = "SO_sim_0"
cnc_params["observables"] = [["q_so_sim"]]
cnc_params["obs_select"] = "q_so_sim"

# Mass and redshift range
cnc_params["M_min"] = 1e13
cnc_params["M_max"] = 1e16
cnc_params["z_min"] = 0.01
cnc_params["z_max"] = 3.0

# Selection observable range
cnc_params["obs_select_min"] = 5.0
cnc_params["obs_select_max"] = 200.0

# Precision parameters
cnc_params["n_points"] = 16384
cnc_params["n_points_data_lik"] = 2048
cnc_params["n_z"] = 100
cnc_params["sigma_mass_prior"] = 10
cnc_params["delta_m_with_ref"] = True
cnc_params["scalrel_type_deriv"] = "numerical"
cnc_params["downsample_hmf_bc"] = 2
cnc_params["compute_abundance_matrix"] = True

# Parallelisation
cnc_params["number_cores_hmf"] = 1
cnc_params["number_cores_abundance"] = 1
cnc_params["number_cores_data"] = 8
cnc_params["parallelise_type"] = "redshift"

# Cosmology parameters
cnc_params["cosmology_tool"] = "classy_sz"
cnc_params["cosmo_param_density"] = "critical"
cnc_params["cosmo_model"] = "lcdm"

# Likelihood parameters
cnc_params["likelihood_type"] = "unbinned"
cnc_params["data_lik_from_abundance"] = True
cnc_params["stacked_likelihood"] = False

# Scaling relation parameters
scal_rel_params["corr_lnq_lnp"] = 0.0
scal_rel_params["bias_sz"] = 0.8

# Initialize cosmocnc
number_counts = cosmocnc.cluster_number_counts()
number_counts.cnc_params = cnc_params
number_counts.scal_rel_params = scal_rel_params
number_counts.cosmo_params = cosmo_params
number_counts.initialise()

# Compute log-likelihood for different bias_sz values
bias_sz_values = np.linspace(0.79, 0.81,40)
log_lik = np.zeros(len(bias_sz_values))

for i, bias_sz in enumerate(bias_sz_values):
    scal_rel_params["bias_sz"] = bias_sz
    number_counts.update_params(cosmo_params, scal_rel_params)
    log_lik[i] = number_counts.get_log_lik()

# Normalize and plot the exponential of the log-likelihood
lik_vec = np.exp(log_lik - np.max(log_lik))
plt.plot(bias_sz_values, lik_vec)
plt.xlabel("Mass bias")
plt.ylabel("Likelihood")
plt.title("Likelihood for Simons Obsevatory galaxy clusters")
plt.axvline(x=0.8,linestyle="dashed",color="k")
plt.ylim(bottom=0.)
plt.savefig("output/results/likelihood_plot_cosmocnc.pdf")
\end{lstlisting}

\end{document}